# Extragalactic Radio Continuum Surveys and the Transformation of Radio Astronomy

Ray P. Norris[1,2]

[1]Western Sydney University, Locked Bag 1797, Penrith South, NSW 1797, Australia
[2]CSIRO Astronomy & Space Science, PO Box 76, Epping, NSW 1710, Australia

**ABSTRACT**
Next-generation radio surveys are about to transform radio astronomy by discovering and studying tens of millions of previously unknown radio sources. These surveys will provide new insights to understand the evolution of galaxies, measuring the evolution of the cosmic star formation rate, and rivalling traditional techniques in the measurement of fundamental cosmological parameters. By observing a new volume of observational parameter space, they are also likely to discover unexpected new phenomena. This review traces the evolution of extragalactic radio continuum surveys from the earliest days of radio astronomy to the present, and identifies the challenges that must be overcome to achieve this transformational change.

## 1. INTRODUCTION

A human observing a clear night sky sees only starlight. Most of this starlight comes from visible stars, including those in distant galaxies, plus objects such as the Moon and planets that are illuminated by our star, the Sun. To an observer with eyes sensitive to radio rather than visible light, stars would be relatively faint. Instead, the sky would be dominated by diffuse radio synchrotron emission from our own Galaxy, plus many pairs of compact objects, typically billions of light years away. These pairs are the twin lobes of radio galaxies, illustrated by Figure 1, caused by a supermassive black hole (SMBH) in the nucleus of some galaxies, termed Active Galactic Nuclei (AGN). Stars and normal galaxies are far weaker at radio wavelengths than these exotic radio galaxies.

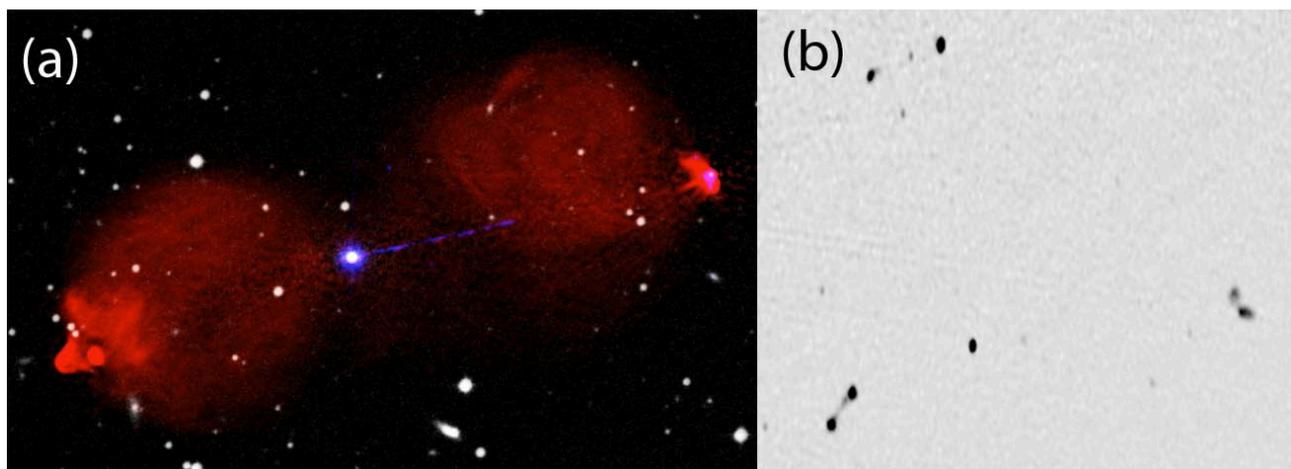

**Figure 1: Typical sources in the radio sky.** *(a): A composite image of a radio galaxy (Pictor A) with radio in red, optical in white, and X-ray in blue. An X-ray jet emanates from the environs of a supermassive black hole at the centre, powering two diffuse lobes (shown in red) of radio emission, which dominate the appearance at radio wavelengths. Image courtesy of Emil Lenc [118]. (b): A sample of a typical deep radio survey [87]. Many of the radio sources are grouped into pairs or triples, which are probably similar to the radio galaxy shown in on the left, but much fainter and more distant.*

The problem facing the first radio astronomers was that these radio galaxies are often very faint at optical wavelengths, so that optical and radio surveys largely sampled two different populations of objects, with few objects in common. A few normal low-redshift star-forming galaxies were found to emit diffuse synchrotron radiation at radio wavelengths analogous to that from our own Galaxy, but this was much weaker than the emission from radio-loud AGN. Of those few AGN that could be identified, most seemed to be either low-redshift elliptical galaxies, or high-redshift radio-loud quasars. It was therefore widely thought that all radio galaxies were elliptical galaxies.

It is now known that there is a continuum of galaxy properties, ranging from pure star-forming galaxies, such as our own Galaxy, whose radio output is dominated by stellar evolution processes, to the radio-loud objects whose radio emission is dominated by the AGN. There is also an important class of composite galaxies in which the radio emission has comparable contributions from the AGN and stellar evolution components.

Fortunately, as the sensitivity of both radio and optical telescopes has increased, the overlap between optical and radio surveys has grown to the point where many optical galaxies are detectable in the radio, and vice versa. Radio surveys are no longer dominated by exotic objects, of little interest to those studying the galaxies seen at optical wavelengths, but are taking their place as a window providing new information on the majority of classes of objects in the sky. Radio surveys are therefore a key tool in understanding the evolution of galaxies over cosmic time.

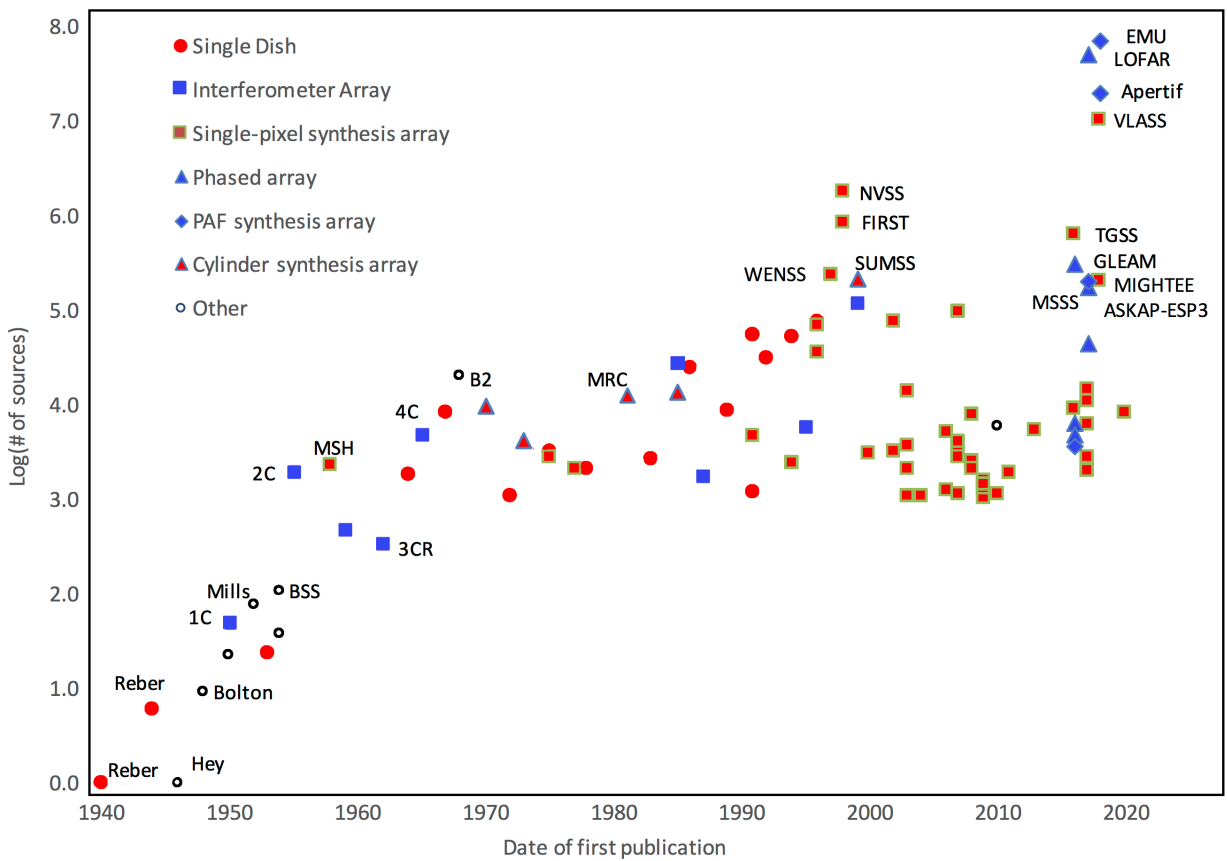

*Figure 2: A plot of the number of known extragalactic radio sources discovered by surveys as a function of time. The increase in the number of radio sources detected by extragalactic radio surveys, from the birth of radio astronomy to the next-generation surveys. Surveys with less than 1000 sources are omitted except for those in the early days of radio astronomy. The symbols indicate the type of telescope used to make the survey as follows: red circle: single dish, blue square: non-synthesis interferometer array, red square: conventional synthesis array, blue triangle: a phased array, blue diamond: a synthesis array using phased-array feeds, red triangle: a*

*cylindrical telescope, open circle: anything else. Details of individual surveys are given in Table 1 (in the online supplementary material).*

This review attempts to chart the changing nature of extragalactic radio continuum surveys, highlight the successes and challenges, and describe the "next generation" radio continuum surveys that will soon dominate the field. For conciseness, this review does not include Galactic, spectral line, or polarisation surveys, nor the experiments to detect and measure the cosmic microwave background or the epoch of reionisation. Figures 2 and 3 compare the performance and historical context of all major extragalactic radio continuum surveys. A list of the most significant radio continuum surveys, which was used to compile these Figures, can be found in the Supplemental Material.

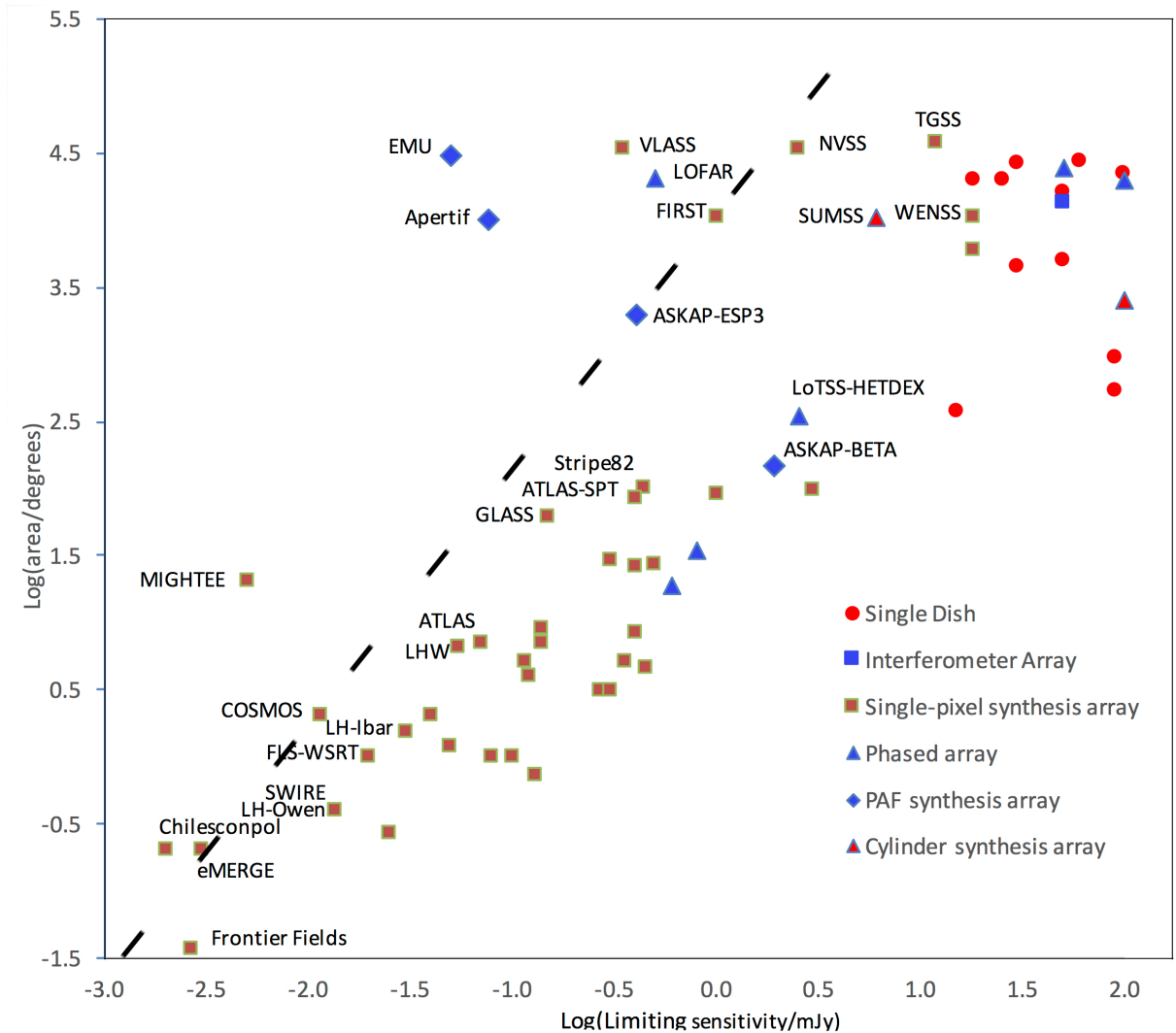

*Figure 3: The sky area vs sensitivity of modern radio surveys. The sensitivity is either the quoted detection limit or 5 times the quoted rms noise level. The dashed line marks the boundary of existing surveys, and roughly corresponds to a few months of observing time on one of the leading international radio telescopes. Symbols are the same as in Figure 2.*

## 2. HISTORY

Radio astronomy started in December 1932 when Karl Jansky [1] found that a component of short-wave "static" came from a fixed position in the sky, which he subsequently found to be the centre of our Galaxy [2]. However, this discovery attracted little attention in the astronomical community. The first systematic surveys of radio sources [3,4,5] quickly found that most of the emission was from the Milky Way. Reber [3] also reported a tentative (but incorrect) detection of M31, and [4] the first true extragalactic source (Cygnus A), which was first detected as a discrete source by Hey et al. [6], and first resolved using the Australian sea-cliff interferometer in 1948 [7]. The first extragalactic identifications of radio sources were made in 1949 [8], although Cygnus A was not identified as an extragalactic source until 1954 [9].

The end of the Second World War saw a surge of activity in radio astronomy, led mainly by ex-radar scientists in Cambridge, Manchester, and Sydney. By 1950 there were 67 known radio sources, although only seven had been identified [10], including M31 [11]. Most of the work focused on measuring their angular size and position, to enable cross-identification. Meanwhile, three groups (Ryle at Cambridge, and Mills and Bolton in Sydney) produced catalogues of sources, to explore their range of properties. By 1954, when Bolton [12] produced the first catalogue with more than one hundred sources, they were still regarded as "Galactic" sources.

In 1950, the strong radio emission from our Galaxy was attributed [13] to synchrotron emission by cosmic rays accelerated by supernova shocks. It was deduced that other normal galaxies should also exhibit this emission, which would be far weaker than the powerful radio galaxies. Few normal galaxies were therefore detected in early surveys.

The first large (>1000 sources) radio survey, published in 1955, was the 2C survey at Cambridge [14] which catalogued 1936 discrete radio sources at 81 MHz. The distribution of flux densities in the 2C catalogue was inconsistent with a Euclidean distribution of standard candles, and was claimed [15] to rule out the "Steady State" theory, favouring instead a "Big Bang" model. However, the 2C survey was subsequently found to contain many spurious sources caused by strong radio sources detected in the outer sidelobes of the telescope point spread function [16]. Nevertheless, the statistical properties of the noise below the detection threshold [17] still favoured the Big Bang model. The first major survey without significant errors, published in 1957-58, was made with the Mills Cross interferometer [18,19] near Sydney, whose 85 MHz catalogue showed a distribution of flux densities only slightly steeper than predicted by a Euclidean model

Subsequent surveys at Cambridge, notably the 3C in 1959 [20], 3CR in 1962 [21] and 4C in 1961[22], agreed with the Sydney surveys, but showed strong evolution of radio sources, from which it was successfully argued that the Steady-State theory was incorrect.

These early successes stimulated the development of radio astronomy groups in The Netherlands, where the Dwingeloo telescope was completed in 1956, the United States, where the Owens Valley Radio Observatory was built in 1958, and Italy, where the construction of the Northern Cross telescope started in 1960, resulting in the Bologna B2 survey [23].

In 1969, the first large single-dish survey with the new Parkes telescope was published [24], and included spectral and polarisation information, as well as optical identifications. Successive single-dish multi-frequency surveys continued with the Parkes telescope for two decades, with a final consolidated catalogue published in 1991 [25]. These surveys resulted in a growing understanding of the properties of AGN, and the physical processes driving them.

The many continuum surveys since then are listed in the Table 1 and shown in Figure 2. From 1990 to 2004, new technology enabled the construction of sensitive new telescopes, resulting in a hundred-fold increase in the total number of known radio sources. This enormous change was dominated by the following four surveys.

1. The Westerbork Northern Sky Survey (WENSS) [26] surveyed a large area of the northern sky between 1991 and 1996 at 327 MHz, producing a catalogue of about 230,000 sources.

2. The NRAO VLA Sky Survey (NVSS) surveyed the northern sky at 1.4 GHz between 1993 and 1996, producing a catalogue of about 1.8 million sources. NVSS is still the largest radio survey, and its survey paper [27] is the second most highly-cited paper in radio astronomy.

3. The complementary "Faint Images of the Radio Sky at Twenty-Centimeters" (FIRST) [28] surveyed a smaller area between 1993 and 2004 with higher resolution and greater sensitivity to yield a catalogue of about 800,000 sources. It found many sources that are not present in the NVSS catalogue, but is insensitive to some extended NVSS sources.

4. The lack of a corresponding southern hemisphere survey was rectified by the Sydney University Molonglo Sky Survey (SUMSS) [29,30] which used the upgraded "Molonglo Cross" telescope to survey the southern sky at 843 MHz from 1997 to 2003, to produce a catalogue of about 211,000 sources.

Figure 2 shows that the size of surveys subsequently plateaued for almost two decades. Surveys during this time focussed on covering smaller areas very deeply, presumably because another large shallow survey could not be justified until the technology enabled an order-of-magnitude improvement over the earlier surveys, particularly NVSS.

**3 THE RADIO SKY**
The twin-lobed radio galaxies that dominated early surveys are caused by the synchrotron radiation from relativistic plasma ejected from material falling towards a SMBH. Sometimes a third component is found between them, at the position of the host, marking the start of the jet of relativistic electrons powering the double lobes. These sources naturally fall into two groups [31]: Fanaroff-Riley Class II (FRII) which are very luminous sources dominated by edge-brightened lobes, and the less luminous FRI sources in which the peak brightness occurs closer to the host.

Orientation can dramatically change the appearance of a radio source and its host, an effect which is sometimes called "the Unified Model" [32]. For example, relativistic beaming in the jets causes sources with jets along the line of sight to have a much brighter nucleus (the so-called core-dominated sources) than those with jets pointed elsewhere [33].
The central broad-line region of most radio galaxies is obscured by dust, but when the jet is oriented close to the line of sight, the broad-line region is visible, giving the galaxy a "stellar" appearance at visible wavelengths, and the radio galaxy is classed as a quasar.

Quasars were originally discovered [36] based on their radio properties, and, by definition, are radio-loud. A corresponding class of radio-quiet objects were called "QSO's" at the time of their discovery [35] but are now normally called "radio-quiet quasars" or RQQ. The difference between radio-quiet and radio-loud objects cannot be explained by orientation, but must be an intrinsic property of the host. The hosts of RQQ are also found to emit weak radio emission, which is sometimes ascribed to the star-formation processes of the host [36,37], but in at least some cases the RQQ shows radio emission from a weak AGN [38,39]. More generally, it is now recognised that there is a class of low-luminosity AGN, typified by radio-quiet quasars and Seyfert galaxies, which mainly appear in radio survey images as faint unresolved sources.

While the strongest sources are AGN, sensitive surveys with sufficient resolution can detect synchrotron emission from normal star-forming galaxies.

These classes of radio source can be seen in Figure 4. At high flux densities, the source counts are dominated by radio-loud AGN, and follow a smooth power law distribution down to about 1 mJy. Below 1 mJy, the source counts flatten, suggesting an additional population of low-luminosity AGN, star-forming galaxies, and composite galaxies. It is remarkably difficult to distinguish between low-luminosity AGN and star-forming galaxies, but there is evidence that the density of star-forming sources approaches that of AGN at a flux density level of about 200 µJy [40].

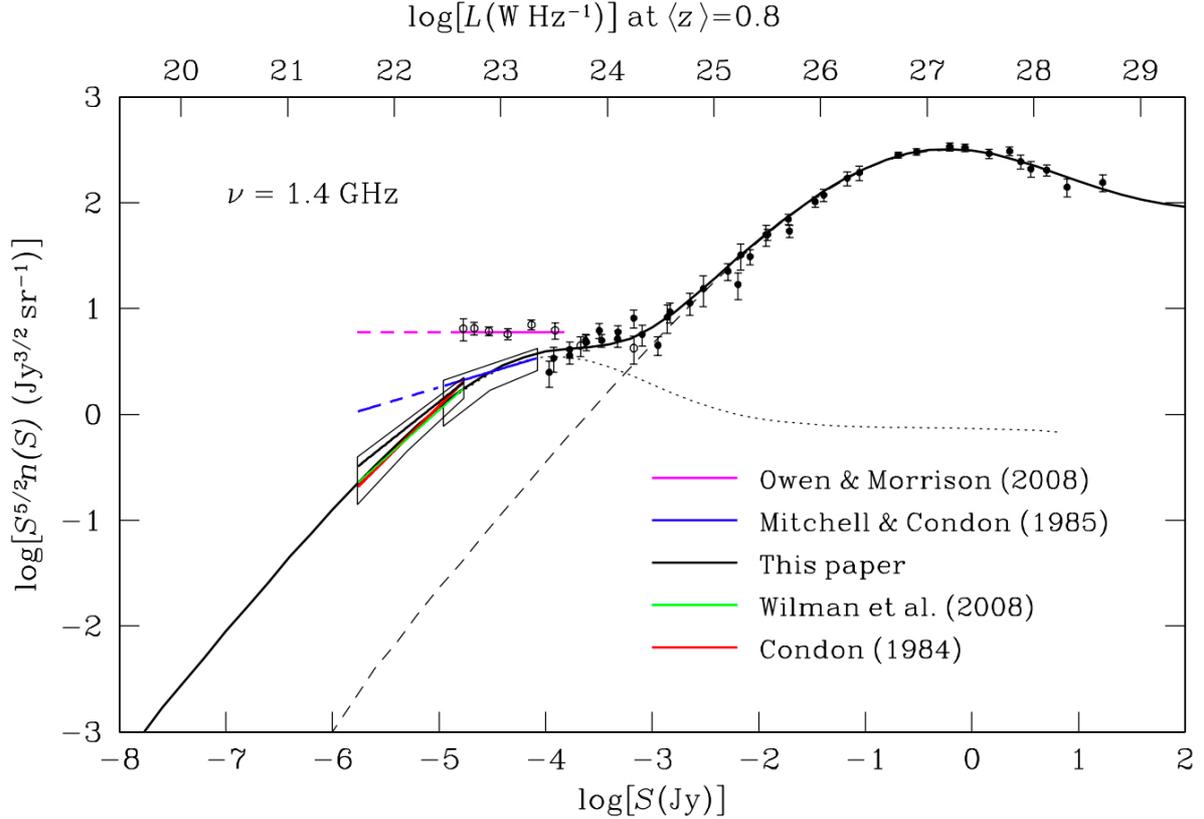

*Figure 4: The number of radio sources as a function of flux density, plotted as a Euclidean-normalised differential source count plot at 1.4 GHz, taken from [85]. The dashed line is the contribution from radio-loud AGN, and the dotted line is the contribution from star-forming galaxies and low-luminosity AGN. Coloured lines show possible extrapolations.*

## 4. WHY DO RADIO SURVEYS?

Radio surveys are an important astrophysical tool to provide large samples of galaxies for studying cosmology or galaxy evolution. These surveys also reveal rare but important stages of galaxy evolution, and expand the volume of observed parameter space, increasing the likelihood of making unexpected discoveries [41]. Here I outline some of the science that drives these surveys. A more detailed discussion can be found elsewhere [40,42,43].

### 4.1 The Evolution of Star Formation

Only a small fraction of radio sources found in the early radio surveys were star-forming galaxies, but increasingly sensitive observations enabled their study at radio wavelengths. Star-forming galaxies are expected to represent about half of the sources in next-generation surveys.

The cosmic star formation rate density (SFRD) of the Universe started at zero, reached a broad maximum about 7 billion years ago, and has since declined by about an order of magnitude to the present day [44]. However, the shape of the curve is poorly defined, particularly at high redshifts, where different star formation indicators give different values [45,46], largely because of poorly-known extinction corrections. The contribution of different classes of galaxy to the overall rate is poorly determined, although it is known that the SFRD at low redshift is dominated by lower-mass galaxies in low-density environments, while the SFRD at high redshifts is dominated by high-mass galaxies in high-density environments that rapidly exhaust their fuel, and are the progenitors of the brightest early-type galaxies at low redshift.

The radio emission from star-forming galaxies is dominated by synchrotron emission generated by cosmic ray electrons that have been accelerated by supernova shocks, and is proportional [47] to the far-infrared (FIR) emission, and therefore to the star-formation rate. This correlation is now known to be accurate over several orders of magnitude [48], and to extend beyond redshift 2 [49]. A plausible explanation [50] is that young stars both heat the dust causing the FIR emission, and explode in supernovae, causing shocks that accelerate the electrons, although this model is unable to explain the observations in detail [51,52]. Nevertheless, measuring the FIR-radio correlation (FRC) is a valuable empirical method for distinguishing between star-forming galaxies and radio-loud AGN.

Since radio waves are unaffected by dust extinction, radio measurements of the synchrotron emission are found to give an accurate measure of star formation rate (SFR). In the future, high-frequency surveys, sensitive to free-free emission, will also be important [53]. While FIR luminosity is also a good measure of SFR, next-generation radio surveys will detect more star forming galaxies per $\deg^2$ than FIR surveys, giving larger sample sizes. Radio continuum surveys therefore provide an important tool for measuring the cosmic star formation history of the Universe. However, radio measurements of SFR need redshifts, and better ways of detecting a contribution from radio-luminous AGN that might corrupt the measurement.

Provided these challenges are met, next-generation radio surveys will measure the evolution of the cosmic star formation rate to an unprecedented accuracy to high redshift, with a large enough sample to examine the contributions from different classes of galaxy.

**4.2 The Evolution of AGN**
One of the major unsolved problems in radio astronomy is to understand the difference between radio-loud and radio-quiet sources [36]. Although low-redshift radio-loud AGN are predominantly hosted by elliptical galaxies [54], the probability of a high-redshift galaxy or quasar possessing a radio-loud AGN seems to depend primarily on the mass of the host [55]. Several factors including black hole spin have been suggested as the cause of this dichotomy [56]. The existence of "restarted" radio galaxies [57] suggests that radio-loudness, or AGN activity, may be an episodic phenomenon, with a quasar spending perhaps 10% of its time in a radio-loud mode, perhaps on timescales as short as $10^5$ years [58].

Simulations of galaxy formation cannot reproduce the observed mass spectrum of galaxies, typically producing too many small galaxies and too few large galaxies. This discrepancy is thought to be caused by feedback mechanisms, in which the energy from supernovae and AGN limit the star formation of a galaxy [59].

Two modes of accretion onto the SMBH are recognised. In cold-mode accretion, which is the dominant accretion mode at high redshift, cold gas fuels a SMBH with high efficiency, but then the resulting AGN activity disrupts the infalling gas, limiting the accretion rate. In hot mode accretion, typically seen in low-redshift elliptical galaxies, the cool gas has either been used up, heated, or expelled from the galaxy, and the remaining hot gas accretes onto the SMBH with low efficiency [60,61]. These mechanisms result in the changing space density [62] and luminosity functions [63] of radio sources. AGN activity peaks in the redshift range 1-2, closely following the evolution of cosmic star formation rate, and suggesting that they are linked, perhaps by AGN feedback.

Future radio surveys will help us understand the links between AGN evolution and star formation evolution, the role and redshift distribution of low-luminosity AGN, and the mechanisms of, and differences between, different classes of radio AGN, and how they are related to the accretion modes.

### 4.3 Clusters of Galaxies

Clusters of galaxies are the most massive bound objects in the Universe, consisting of galaxies in a cloud of intergalactic gas, located at the intersections of filaments and sheets of the cosmic web. Clusters are rich in dynamics, shocks, interactions, and, at high redshift, even molecular gas [64]. The radio emission from clusters consists of three elements: (a) the radio emission from constituent galaxies, including bent-tail (BT) galaxies that are interacting with the intracluster gas, (b) diffuse elongated objects, known as "relics", caused by shock-excited electrons [65], and (c) a large diffuse "halo" sometimes found centred on the cluster core. Our understanding of these has been hampered by insufficient radio data. For example, most of the ~60 known radio haloes [66,67] have been found in clusters detected first at X-ray wavelengths, and our knowledge may be biased by selection effects. Next-generation radio surveys will discover hundreds of diffuse radio haloes and thousands of BT galaxies. Radio emission in clusters is therefore likely to become an important tool both for studying clusters themselves, and for detecting large numbers of clusters to study cosmology and trace large-scale structure formation.

### 4.4 Cosmology

Since the early use of surveys to argue against the Steady State theory, there have been many attempts to use radio surveys to measure cosmological parameters, with some successes such as the measurement of the cosmic dipole [68]. However, only now are radio continuum surveys approaching the size and depth necessary to rival other wavelengths at the precision measurement of cosmological parameters. One technique, to measure the distortion of the images of radio sources by weak gravitational lensing [69], will probably not be feasible until the advent of large surveys with the Square Kilometre Array. Other techniques [70] use three indicators of the statistical properties of catalogues of radio surveys, via the angular power spectrum, cosmic magnification, and the Integrated Sachs-Wolfe effect. Even if no redshifts are available, next-generation surveys will yield independent measurements of cosmological parameters [70] to complement those from dedicated projects such as Euclid [71] and Dark Energy Survey [72]. However, even without accurate individual redshifts, statistical redshifts [40], in which objects are assigned to a small number of redshift bins, can significantly increase the accuracy of radio-derived cosmological parameters, making them likely to become important cosmological tools [73].

### 4.5 Discovering the Unexpected

While science resulting from well-defined science goals is important, experience shows that most major discoveries in astronomy are unexpected, often when technical innovation enables a new part of the observational parameter space to be observed [74]. Figure 3 shows that next-generation radio surveys explore a new part of parameter space, and so should yield unexpected discoveries, provided they are equipped to do so. However, the complexity of the telescopes and the large data volumes mean that it may be difficult for a human to make these discoveries. Instead, most science will be extracted from the large survey datasets by querying the data with specific questions, resulting in specific answers. To discover the unexpected, we need to develop algorithms [41] that can mine the data for the unexpected, such as that developed by [75] to identify "weird" galaxies in SDSS data by looking for abnormal spectra.

## 5. TECHNICAL CHALLENGES TO RADIO SURVEYS

Technical innovation is a major driver of radio astronomy, and major discoveries often follow the adoption of a new technology [74]. While technology continues to increase the sensitivity and bandwidth of telescopes, new analysis techniques such as compressive sampling [76] and machine learning [77] compete with traditional techniques.

Perhaps the greatest challenge facing next-generation radio telescopes is radio frequency interference (RFI). It is best minimised by locating the telescopes in a radio-quiet site, such as those chosen for the Square Kilometre Array [78] in Australia and South Africa. However, these sites, although orders of magnitude better than urban sites, are still susceptible to satellites, aircraft, and even terrestrial interference reflected from the Moon [79], which has itself been suggested as a future radio observatory site [80]. A promising avenue is the development of active interference mitigation techniques, such as the use of null beams that track an interfering signal for subsequent subtraction from the data [81].

Early radio continuum surveys chose low frequencies because of technical limitations, and because most radio sources are stronger at low frequency, although RFI is often worse at low frequencies too. Higher frequencies offer a higher resolution and positional accuracy, and higher dynamic range. These factors determine a "sweet spot" of about 1-3 GHz chosen by most current large radio continuum surveys, although MWA [82] and LOFAR [83] use low frequencies, and AT20G [84] uses high frequencies, to optimise the detection of different physical mechanisms.

Confusion, in which the beam size fails to distinguish between neighbouring sources, represents a fundamental limit to surveys. For example, a sea of faint unresolved sources limits the sensitivity of contemporary 20 cm surveys to a few µJy/beam [85], although techniques are being explored [86] to use the statistical properties of noise to probe the astrophysics of faint sources. Confusion caused by strong sources in the sidelobes of the telescope may be overcome by techniques such as constructing an a priori model of strong sources in the radio sky which is subtracted from the observed data before imaging.

Cross-matching radio sources with optical/infrared catalogues is essential but difficult. About 90% [87] of sub-mJy radio sources are simple unresolved sources at arcsec resolution, and can be reliably matched using traditional techniques such as the Likelihood Ratio [88]. However, the remaining 10% are complex sources consisting of several radio components. For example, two nearby unresolved radio components might either be the two lobes of an FRII radio source, in which case the optical host lies near their midpoint, or they might be the radio emission from two unassociated star-forming galaxies. Distinguishing between these two cases is difficult, but techniques are being developed to address this challenge, including estimating the probabilities of competing hypotheses about a particular source [89], or training a convolutional neural network on a set of sources which have been classified manually [41,90], or using the classification abilities of thousands of citizen scientists in projects such as Radio Galaxy Zoo [91].

Redshifts are important for many of the science goals of radio continuum surveys. In most cases, redshifts will be obtained from optical/IR surveys, or from HI surveys such as WALLABY [92], LADUMA [93], and APERTIF [94]. Spectroscopic redshifts are currently impractical for surveys of tens of millions of sources, so photometric redshifts must be used. Because of strong AGN evolution, traditional template-based photometric redshifts are not well-matched to the high-redshift AGN found in radio surveys, nor exploit the available radio photometry. Empirical methods, such as machine learning, have the advantage of being able to use other available data, such as radio morphology and polarization [95], and can use training sets from deep multiwavelength surveys. Furthermore, the wide fractional bandwidth of modern surveys has shown that a source's radio spectrum is often not a simple featureless power law, but has an intrinsic rest-frame spectral energy distribution containing features [96, 97, 98] which may eventually enable the estimation of redshifts using radio photometric data alone.

6.     NEXT-GENERATION CONTINUUM SURVEYS
Since the transformational period in 1990-2000 when WENSS, NVSS, FIRST, and SUMSS increased the number of known radio sources from tens of thousands to about 2.5 million, there have been many important deep surveys, but none have had the transformational power of those four. However, radio astronomy is now embarking upon another transformational period, with the advent of seven major continuum surveys, each based on a new, or radically upgraded, telescope. These surveys not only have enormously increased sensitivity, but are also transformational in their polarisation capability and bandwidth. Over the next few years, these surveys will increase the number of known radio sources by a factor of about 40. The areas of sky where they overlap with each other or with other multiwavelength surveys, will be particularly important scientifically.

The Australian SKA Pathfinder (ASKAP) [99] is a new radio telescope approaching completion in Western Australia with a maximum baseline of 6 km, operating at 700 to 1800 MHz. Each of the 36 antennas is equipped with a Phased Array Feed (PAF) [100] giving it a field of view of 30 deg$^2$, resulting in a high survey speed. The telescope is currently in an "early science" phase [101], with full operations expected to start in early 2018. ASKAP's continuum survey is the Evolutionary Map of the Universe (EMU) [102] which plans to survey the entire visible sky to an rms sensitivity of 10 μJy/beam. EMU is expected to generate a catalogue of about 70 million galaxies at 1100 MHz, with spectral shapes and all polarisation products (courtesy of the POSSUM project [103]) across a 300 MHz band.

The Giant Metrewave Radio Telescope (GMRT) [104] is a synthesis array in India with 30 antennas spanning 25 km, operating over the frequency range 150 to 1500 MHz. Formally opened in 2001, it is currently undergoing a major upgrade to extend its sensitivity, reliability, and frequency coverage. A large all-sky survey at 150 MHz called the TIFR GMRT Sky Survey (TGSS) was started in 2009. However, processing the data has proved more challenging than expected, and only a small fraction of the survey has been published. A re-analysis of the raw archived data [105] has yielded a catalogue of 0.63 million sources, reaching an rms noise below 5 mJy/beam. The upgraded GMRT is expected to host even larger surveys.

LOFAR, the Low Frequency Radio Array [83], is a newly-completed phased-array telescope, operating at 10 to 240 MHz, that spans an area of 100 km diameter in the Netherlands, with additional stations located in Germany, UK, Sweden, France, and Poland. Each station is an array of antennas that form many beams simultaneously on the sky. The Multifrequency Snapshot Sky Survey (MSSS) [106] first used LOFAR to make a shallow first-pass of the northern sky, cataloguing over 150,000 sources. The main LOFAR continuum survey, LoTSS, [107,108,109] consists of three tiers: Tier 1 is a shallow wide-field survey that is expected to detect 50 million radio sources. Tier 2 is a deep survey over 500 deg$^2$ targeted at deep fields, clusters, and nearby galaxies. Tier 3 includes 5 single deep pointings at 150 MHz to reach the confusion level of 5 μJy/beam rms.

MeerKAT [110], the South African SKA pathfinder telescope, is nearing completion in the Karoo region. Its 64 antennas span an area 8 km in diameter. MeerKAT continuum surveys are still being planned, but will probably include the MIGHTEE survey [111] which will survey about 20 deg$^2$ at about 1.4 GHz to the confusion level of about 1 μJy/beam, detecting about 200,000 sources. Polarized sources should be detectable significantly below this, and statistical techniques may provide astrophysical information at even deeper levels [86].

The Murchison Widefield Array (MWA) [82] is a low-frequency synthesis telescope in Western Australia, located adjacent to ASKAP. It currently consists of 128 'tiles' of dipoles operating in the 80-300 MHz frequency range, over an area of 3 km in diameter. It is currently being upgraded to longer baselines and more tiles. The main continuum survey is GLEAM [112] which has produced a catalogue [113] containing over 300,000 sources at a typical rms of 9 mJy/beam. Further data releases will reach lower flux densities with higher resolution and will include the Galactic Plane and other areas omitted from the initial catalogue.

The Very Large Array (VLA) is a versatile 27-antenna synthesis array with a maximum baseline of 35 km operating in the frequency range 0.074 to 50 GHz, and has been in operation since 1980. It has been responsible for several of the leading radio surveys shown in Figure 3, including the NVSS which is currently the largest radio survey. It was upgraded in 2011 to become the Karl G. Jansky Very Large Array [114] with a major increase in sensitivity, bandwidth, and operating flexibility. A new radio continuum survey has started, called the VLASS [115], which will survey from declination -40 to +90° at 2-4 GHz. The survey is expected to reach an rms of 70 μJy/beam in 2023, producing a catalogue of about 10 million sources.

The Westerbork Synthesis Radio Telescope (WSRT) is a 14-antenna array with a maximum baseline of 2.7 km, which has been in operation since 1970. The recent APERTIF upgrade [94] installed PAF receiver systems operating over a frequency range of 1.0 to 1.7 GHz. The APERTIF continuum surveys are still being planned, but will probably include the WODAN survey [116] which will survey the northern cap of the sky that is inaccessible to ASKAP, to a target rms sensitivity of 15 μJy/beam. It is also planned to observe a deeper tier to a target rms of about 5 μJy/beam to cover 20 deg$^2$ being surveyed by the LOFAR Tier 3 survey.

## 7. CONCLUSION: THE FUTURE

Radio continuum surveys have a proud history of generating major breakthroughs in our understanding of the Universe. The next-generation surveys now starting will probe unexplored areas of observational parameter space, and history suggests that we can expect revolutionary discoveries as a result.

These surveys will also change the nature of radio astronomy. Existing radio measurements are not as intrinsically deep as existing optical data, and so >99% of objects studied at optical wavelengths have no radio data. Next-generation surveys are crossing a sensitivity threshold below which most galaxies detected in radio surveys are normal star-forming galaxies, and many galaxies found in optical/IR surveys will have radio photometry. About 20% of galaxies detected by surveys such as SDSS and WISE will be detected by the new radio surveys, and radio astronomical measurements will become an indispensable part of every astronomer's toolkit.

The numbers of galaxies detected in these surveys are increasing to tens of millions, suggesting that new approaches will be needed to generate the science from the data. These surveys are likely to differ from the previous generation in three ways: (a) Next-generation surveys will routinely generate polarisation and spectral shape measurements that were previously available for only a few radio sources, (b) Techniques for extracting the science from the data will change, with an increasing emphasis on empirical techniques such as machine learning; (c) The approaches for interpreting the science from the data will change from studying features of individual galaxies to studying the statistical properties of subsamples. We can view the surveys as sampling the many stages, and many byways, of the evolutionary paths of galaxies from soon after the Big Bang through to the present day. The challenge will be to identify the several evolutionary threads, and place the surveyed galaxies in their time and place on this web of evolutionary sequences.

This review has said little about the Square Kilometre Array (SKA) [78], whose greatest strength will be to make extremely deep observations over small areas of sky. Phase 1 of the SKA, which is scheduled to be completed about 2023, may conduct even larger surveys than those discussed here, although specific plans are still being debated. A decade further on will hopefully see the construction of SKA Phase 2, which is planned to outperform current telescopes by at least an order of magnitude, and will revolutionise radio astronomy yet again.

Radio continuum surveys are at a watershed. Behind us are the first tentative steps of discovery, then the gradual realization of the vast diversity of radio sources. Ahead of us is the data-driven era in which we apply ingenuity to devising the key questions with which to mine our samples of tens of millions of objects. Radio astronomy is about to take its place in the toolbox of every astronomer, opening a new window of radio photometry on many objects studied at other wavelengths. Above all, we are opening up new tracts of unexplored parameter space, in which history tells us we are likely to make completely unexpected discoveries, provided we have the tools and the insight to do so.

**AUTHOR INFORMATION**
Correspondence and requests for materials should be addressed to Ray Norris at R.Norris@westernsydney.edu.au

**ACKNOWLEDGEMENTS**
Some of the information in Table 1 was taken from tables kindly shared by Heinz Andernach[117], Isabella Prandoni, and Joe Callingham. I thank the following for contributing to or commenting on an early draft of this review: Heinz Andernach, Joe Callingham, Claire Chandler, Jim Condon, Erwin de Blok, Ron Ekers, Miroslav Filipovic, Chris Hales, George Heald, Natasha Hurley-Walker, Amy Kimball, Roland Kothes, Mark Lacy, Emil Lenc, Tom Muxlow, Eric Murphy, Tom Oosterloo, Isabella Prandoni, Huub Röttgering, Nick Seymour, Vernesa Smolcic, Russ Taylor, Randall Wayth.

**Competing interests**
The author declares no competing financial interests.